\documentclass{article}
\usepackage{amsmath,amssymb}

\usepackage{graphicx}

\newtheorem{prop}{Proposition}
\newtheorem{lem}{Lemma}

\title{Ultradiscrete limit of Bessel function type solutions of the Painlev\'{e} III equation}
\author{Shin Isojima}
\date{}
\begin{document}
\maketitle
\begin{center}
Department of Industrial and System Engineering, \\
Hosei University, \\
3-7-2 Kajino-cho, Koganei-shi, Tokyo 184-8584, Japan
\end{center}
E-mail: isojima@hosei.ac.jp 
\section*{Abstract}
An ultradiscrete analog of the Bessel function is constructed by taking the ultradiscrete limit for a $q$-difference analog of the Bessel function. Then, a direct relationship between a class of special solutions for the ultradiscrete Painlev\'{e} III equation and those of the discrete Painlev\'{e} III equation which have a determinantal structure is established. 
\section{Introduction}
Ultradiscretization\cite{TTMS} is a limiting procedure to reduce a given difference equation into a piecewise linear equation, which is written by addition, subtraction and $\max$ operation among dependent variables. In this procedure, we first replace a dependent variable $x_n$ in a given difference equation into $X_n$ with 
\begin{equation}
x_n = e^{\frac{X_n}{\varepsilon}}, \label{eq:exp}
\end{equation}
where $\varepsilon > 0$ is a parameter. Then, applying $\varepsilon\log$ to both side of the equation and then taking the limit $\varepsilon \to +0$. Noting the exponential laws and the identity
\begin{equation}
\lim_{\varepsilon \to +0} \varepsilon\log \left(e^{\frac{X}{\varepsilon}} + e^{\frac{Y}{\varepsilon}}\right) = \max (X,Y),
\end{equation}
multiplication, division and addition among $x_n$ are replaced with addition, subtraction and $\max$ operation, respectively. 
The resulting piecewise linear equation can be regarded as time evolution rule of a cellular automaton since its dependent variables will take discrete values as well as independent variables. 
Moreover, a solution of ultradiscrete system is automatically constructed if we have a exact solution of the original equation. The box and ball system \cite{TS}, which is an ultradiscrete analog of the KdV equation, is well known as one of the most famous examples of ultradiscretization. The soliton solutions of the box and ball system are obtained from those of the discrete KdV equation through the ultradiscrete limit.

However, $x_n$ must be positive if we apply \eqref{eq:exp}. An extended procedure \cite{MIMS09}, \textit{ultradiscretization with parity variables} (p-ultradiscretization), is proposed to overcome this restriction. As an example, we consider a simple equation
\begin{equation}
x_{n+1} = -a x_n,\quad x_0 >0, a>0. \label{eq:ex1}
\end{equation}
Since its solution 
\begin{equation}
x_n = x_0 (-a)^{n}  \label{eq:ex1-1}
\end{equation}
has indefinite sign, it is not possible to ultradiscretize this equation. We introduce the parity (or sign) variable $\xi_n = x_n/|x_n| \in \{1, -1\}$ and define a function
\begin{align}
s(\xi)=
\begin{cases}
1 & \xi=1 \\
0 & \xi=-1.
\end{cases}
\end{align}
Note that a sign $\xi$ is represented by $\xi = s(\xi) - s(-\xi)$. We also introduce the amplitude variable for $x_n$ by $|x_n|=e^{\frac{X_n}{\varepsilon}}$. Then, we rewrite $x_n$ as
\begin{align}
x_n = \{ s(\xi_n) - s(-\xi_n)\} e^{\frac{X_n}{\varepsilon}} \label{eq:s-exp}
\end{align}
and employ \eqref{eq:s-exp} instead of \eqref{eq:exp}. Substituting \eqref{eq:s-exp} and $a=e^{\frac{A}{\varepsilon}}$ into \eqref{eq:ex1} and transposing the negative terms into the other side of the equation, we have
\begin{align}
s(\xi_{n+1})e^{\frac{X_{n+1}}{\varepsilon}} + s(\xi_{n})e^{\frac{X_{n}+A}{\varepsilon}}
= s(-\xi_{n+1})e^{\frac{X_{n+1}}{\varepsilon}} + s(-\xi_{n})e^{\frac{X_{n}+A}{\varepsilon}}. \label{eq:ex1-2}
\end{align}
We apply $\varepsilon\log$ to both side of the equation and then take the limit $\varepsilon \to +0$. If we define a function $S$ by
\begin{equation}
S(\xi) = 
\begin{cases}
0 & \xi=1 \\
-\infty & \xi=-1,
\end{cases}
\end{equation}
the identity
\begin{equation}
\lim_{\varepsilon \to +0} \varepsilon\log \left(s(\xi)e^{\frac{X}{\varepsilon}} + e^{\frac{Y}{\varepsilon}}\right) = \max (S(\xi)+X,Y)
\end{equation}
holds, where the term including $-\infty$ vanishes from $\max$. We may regard $S(\xi)$ as formal ultradiscrete analog of $s(\xi)$ with $s(\xi) = e^{\frac{S(\xi)}{\varepsilon}}$. Utilizing this identity, \eqref{eq:ex1-2} is ultradiscretized as
\begin{align}
&\max( S(\xi_{n+1})+X_{n+1}, S(\xi_{n}) + X_{n}+A)\nonumber \\
{}&= \max (S(-\xi_{n+1}) + X_{n+1}, S(-\xi_{n})+X_{n}+A). \label{eq:ex1-3}
\end{align}
If $\xi_n=1$, this equation turns into $\max( S(\xi_{n+1})+X_{n+1}, X_{n}+A)= S(-\xi_{n+1}) + X_{n+1}$ and it is uniquely solved by $(\xi_{n+1}, X_{n+1})=(-1,X_{n}+A)$. Considering all other cases, \eqref{eq:ex1-3} is rewritten into the explicit form
\begin{align}
\begin{cases}
\xi_{n+1} =-\xi_n \\
X_{n+1} = X_n +A
\end{cases}
\end{align}
and its solution is
\begin{align}
\xi_n &= \xi_0 (-1)^{n} \label{eq:ex1-4}\\
X_n &= nA + X_0. \label{eq:ex1-5}
\end{align}
Here, we note that the sign and order of \eqref{eq:ex1-1} correspond to \eqref{eq:ex1-4} and \eqref{eq:ex1-5}, respectively. The following lemma \cite{IST} provides a useful sufficient condition to obtain a p-ultradiscrete solution. 
\begin{lem}
If a solution of a given difference equation $x_n(\varepsilon)$, where $\varepsilon$ is an arbitrary positive parameter, is evaluated as
\begin{align}
x_n(\varepsilon) = (-1)^{\hat{\xi}_n}e^{\frac{X_n}{\varepsilon}}(c_n + O(\varepsilon)),\quad c_n>0, \label{eq:suff_evaluation}
\end{align}
a pair of the sign $(-1)^{\hat{\xi}_n}$ and amplitude $X_n$ solve the corresponding p-ultradiscrete equation. $\blacksquare$
\end{lem}

An application of p-ultradiscretization to the Painlev\'{e} II equation is reported recently. A $q$-difference analog of the Painlev\'{e} II equation possesses a class of special solutions written in terms of the determinant whose entries are given by solutions of the $q$-Airy equation \cite{HKW}. Solutions of initial value problems to the p-ultradiscrete Painlev\'{e} II equation are presented in \cite{IKMMS} and \cite{IS}, and the direct relationship between these ultradiscrete solutions and discrete solutions is clarified in \cite{IST} through the procedure of p-ultradiscretization. In addition, the p-ultradiscrete analog of the  Painlev\'{e} VI equation and its special solutions are presented and asymptotic behavior of solutions is discussed in \cite{TT}. It is an interesting problem to construct p-ultradiscrete analogs of other Painlev\'{e} equations and study their solutions as fundamental examples of ultradiscrete system with parity variables. 

In this paper, we focus on a discrete analog of the Painlev\'{e} III equation ($\mathrm{dP_{III}}$) which is written as \cite{KOS95}
\begin{align}
w(n+1) w(n-1) = \frac{\alpha {w(n)}^2 + \beta \lambda^n w(n) + \gamma \lambda^{2n}}{{w(n)}^2 + \delta w(n) + \alpha}, \label{eq:qPIII}
\end{align}
where $\alpha$, $\beta$, $\gamma$, $\delta$ and $\lambda$ are parameters. If we take a special set of parameters
\begin{gather}
\alpha=-q^{4N}, \quad \beta = (q^{\nu+N}-q^{-\nu-N-2})q^{8N}(1-q)^2, \quad\nonumber \\
 \gamma=q^{2(6N-1)}(1-q)^4, \quad \delta=(q^{\nu-N}-q^{-\nu+N})q^{2N}, \quad\lambda=q^2, \label{eq:sp_params.}
\end{gather}
\eqref{eq:qPIII} possesses a class of special solutions as follows. A $q$-difference analog of the Bessel equation is written as
\begin{equation}
J_\nu(q^2x)-(q^\nu +q^{-\nu})J_\nu(qx)+\{1+(1-q)^2x^2\}J_\nu(x)=0, \label{eq:qBeq}
\end{equation}
where $q$ is a multiplicative difference interval which satisfies $|q|<1$. We write 
\begin{align}
J_\nu(q^n)=\mathrm{J}_\nu (n) \label{eq:rmJ}
\end{align}
and consider a function with determinantal structure, 
\begin{align}
\tau_N^\nu(n) = 
\begin{vmatrix}
\mathrm{J}_\nu(n) & \mathrm{J}_\nu(n+1) & \cdots & \mathrm{J}_\nu(n+N-1) \\
\mathrm{J}_\nu(n+2) & \mathrm{J}_\nu(n+3) & \cdots & \mathrm{J}_\nu(n+N+1) \\
\vdots & \vdots & \ddots & \vdots \\
\mathrm{J}_\nu(n+2N-2) & \mathrm{J}_\nu(n+2N-1) & \cdots & \mathrm{J}_\nu(n+3N-3)
\end{vmatrix}.\label{eq:tau_det}
\end{align}
Then, functions defined by
\begin{align}
w_N^\nu(n) = \frac{\tau_{N+1}^{\nu}(n+1) \tau_{N}^{\nu+1}(n)}{\tau_{N+1}^{\nu}(n) \tau_{N}^{\nu+1}(n+1)}-q^{\nu+N}
\end{align}
solves \eqref{eq:qPIII} with parameters \eqref{eq:sp_params.}. Note that \eqref{eq:tau_det} satisfies bilinear equations 
\begin{align}
\tau^\nu_{N+1}(n)\tau^{\nu+1}_N(n+1)&-q^{-\nu-N}\tau^\nu_{N+1}(n+1)\tau^{\nu+1}_N(n) \nonumber\\
&=-(1-q)q^{n+2N}\tau^{\nu+1}_{N+1}(n)\tau^\nu_N(n+1),  \label{eq:bilinear1} \\[5pt]
\tau^{\nu+1}_{N+1}(n)\tau^\nu_N(n+1)&-q^{\nu-N+1}\tau^{\nu+1}_{N+1}(n+1)\tau^\nu_N(n) \nonumber\\
&=(1-q)q^{n+2N}\tau^\nu_{N+1}(n)\tau^{\nu+1}_N(n+1),  \label{eq:bilinear2} \\[5pt]
\tau^\nu_{N+1}(n)\tau^{\nu+1}_N(n+3)&-q^{-\nu-N}\tau^\nu_{N+1}(n+1)\tau^{\nu+1}_N(n+2) \nonumber\\
&=-(1-q)q^n\tau^{\nu+1}_{N+1}(n)\tau^\nu_N(n+3),  \label{eq:bilinear3} \\[5pt]
\tau^{\nu+1}_{N+1}(n)\tau^\nu_N(n+3)&-q^{\nu-N+1}\tau^{\nu+1}_{N+1}(n+1)\tau^\nu_N(n+2) \nonumber\\
&=(1-q)q^n\tau^\nu_{N+1}(n)\tau^{\nu+1}_N(n+3)  \label{eq:bilinear4} 
.
\end{align}

Our aim is to derive a class of special solutions for ultradiscrete analogs of \eqref{eq:bilinear1}--\eqref{eq:bilinear4} from \eqref{eq:tau_det} through the ultradiscrete limit. It is unavoidable to introduce p-ultradiscretization for the purpose since a solution of \eqref{eq:qBeq} shows oscillating behavior and the determinant \eqref{eq:tau_det} itself includes negative terms. This paper is organized as follows. In section 2, we evaluate a $q$-difference analog of the Bessel function preparatory for main contents. Based on this result, we evaluate \eqref{eq:tau_det} into which we substitute the $q$-difference Bessel function in section 3. Finally, concluding remarks are given in section 4.
\section{$q$-Bessel function and its ultradiscrete limit}
In this section, we study a special solution of \eqref{eq:qBeq} with $\nu \in \mathbb{Z}_{\ge0}$, which we refer to as the \textit{$q$-Bessel function} in this paper. It is given by
\begin{equation}
J_\nu(x)=(1-q)^{\nu}x^\nu\sum_{j=0}^{\infty}\frac{(-1)^j(1-q)^{2j}}{(q^2;q^2)_{j}(q^2;q^2)_{\nu+j}}x^{2j}, \label{eq:q-besselfnc}
\end{equation}
where
\begin{equation}
(a;q)_k=
\begin{cases}
1 & (k=0) \\
(1-a)(1-aq)\dots(1-aq^{k-1}) & (k\in\mathbb{Z}_{>0}).
\end{cases}
\end{equation}
Jackson's $q$-Bessel function, which is defined even for $\nu \in \mathbb{C}$,
\begin{equation}
J_{\nu}^{(1)}(x;q)=\frac{(q^{\nu+1};q)_{\infty}}{(q;q)_{\infty}}\left(\frac{x}{2}\right)^{\nu}{{}_2\phi_1}\left({{0,0}\atop{q^{\nu+1}}};q,-\frac{x^2}{4}\right) \label{eq:Jackson1}
\end{equation}
is well known, where ${}_r\phi_s$ is the basic hypergeometric series
\begin{equation}
{{}_r\phi_s}\left({{a_1,\dots,a_r}\atop{b_1,\dots,b_s}};q,z\right)=\sum_{j=0}^{\infty}\frac{(a_1;q)_j\dots(a_r;q)_j}{(q;q)_j(b_1;q)_j\dots(b_s;q)_j}[(-1)^jq^{\frac{j(j-1)}{2}}]^{1-r+s}z^j.
\end{equation}
The relationship between \eqref{eq:q-besselfnc} and \eqref{eq:Jackson1}  for $\nu \in \mathbb{Z}_{\ge0}$
\begin{equation}
J_\nu(x) = J_{\nu}^{(1)}(2(1-q)x;q^2) \label{eq:JacksontoqB}
\end{equation}
is readily found.
\subsection{Evaluation of the $q$-Bessel function}
Our aim in this subsection is to prove the following proposition.
\begin{prop}\label{prop2.1}
(i) The $q$-Bessel function $J_\nu(x)$ is deformed as
\begin{equation}
J_\nu(x)=\frac{(1-q)^\nu x^\nu}{(-(1-q)^2x^2;q^2)_\infty}\sum_{k=0}^{\infty}\frac{(-1)^kq^{2k(k+\nu)}(1-q)^{2k}}{(q^2;q^2)_k(q^2;q^2)_{k+\nu}}x^{2k}.\label{eq:qEqBessel}
\end{equation}
(ii) The evaluation as $q\to 0$ 
\begin{align}
J_\nu(q^n)=
\begin{cases}
q^{n\nu}\left(1+O(q)\right)&(n\ge 1)\\
q^{n(n+\nu-1)}\left(\frac{1}{2}+O(q)\right)&(0\ge n\ge -\nu)\\
(-1)^{\frac{n+\nu}{2}}q^{\frac{n(n-2)-\nu^2}{2}}\left(\frac{1}{2}+O(q)\right)&(n\le -\nu-1, n+\nu:\mbox{\rm even})\\
(-1)^{\frac{n+\nu+1}{2}}q^{\frac{n(n-2)-\nu^2+3}{2}}\left(1+O(q)\right)&(n\le -\nu-1, n+\nu:\mbox{\rm odd})
\end{cases}\label{eq:evaluation}
\end{align}
holds for $n\in\mathbb{Z}$. $\blacksquare$
\end{prop}

In order to prove Proposition \ref{prop2.1} (i), we introduce well-known formula \cite{GR}
\begin{align}
\frac{1}{(x;q)_{\infty}}&=\sum_{k=0}^{\infty}\frac{1}{(q;q)_k}x^k \label{eq:basicexp}\\
(x;q)_{\infty}&=\sum_{k=0}^{\infty}\frac{(-1)^kq^{\frac{k(k-1)}{2}}}{(q;q)_k}x^k \\
(a;q)_{\nu+k}&=(a;q)_{\nu}(aq^{\nu};q)_k
\end{align}
and the following lemma.
\begin{lem}
(i) The `\textit{$q$-Euler transformation}' \cite{Exton}
\begin{equation}
\sum_{j=0}^{\infty}c_jd_jx^j=\sum_{k=0}^{\infty}\frac{(\hat{D}^kc_0)}{[k]!}x^k\hat{B}^kf(x)
\end{equation}
holds, where
\begin{align}
f(x) &:= \sum_{j=0}^{\infty} d_j x^j \label{eq:psi_lem1} \\
\hat{B} f(x) &:= \frac{f(x) - f(qx)}{(1-q)x} \\
[k] &:= \frac{1-q^k}{1-q}, \quad [k]! := 
\begin{cases}
1 & (k=0) \\
[k] [k-1] \dots [1] & (k \in \mathbb{Z}_{>0})
\end{cases} \\
\hat{E} c_j &:= c_{j+1}\ (j \in \mathbb{Z}_{\ge 0}), \quad \hat{D}^k := (\hat{E}-1)(\hat{E}-q) \dots (\hat{E}-q^{k-1}).
\end{align}
(ii) A formula 
\begin{align} \label{eq:Dka0}
\hat{D}^k c_0 &= \sum_{j=0}^{k} (-1)^j q^{\frac{j(j-1)}{2}} 
\begin{bmatrix}
k \\ j
\end{bmatrix}
c_{k-j}
\end{align}
holds, where
\begin{align}
\begin{bmatrix}
k \\ j
\end{bmatrix}
&:= \frac{[k]!}{[k-j]! [j]!}=\frac{(q;q)_k}{(q;q)_{k-j} (q;q)_j}. 
\end{align}
$\blacksquare$
\end{lem}
(Proof of Proposition \ref{prop2.1} (i)): 
We consider a function
\begin{equation}
\hat{J}_{\nu}^{(1)}(t;q)=\sum_{j=0}^{\infty}\frac{(-1)^j}{(q;q)_{j+\nu}(q;q)_j}t^j. \label{eq:tempJ}
\end{equation}
Note that 
\begin{equation}
\left(\frac{x}{2}\right)^{\nu}\hat{J}_{\nu}^{(1)}((x/2)^2;q)=J_{\nu}^{(1)}(x;q). \label{eq:temptoJackson}
\end{equation}

If we put 
\begin{align}
&c_j=\frac{1}{(q;q)_{\nu+j}},&&d_j=\frac{(-1)^j}{(q;q)_j},&
\end{align}
we have $\hat{J}_{\nu}^{(1)}(t;q)=\sum c_jd_jt^j$. Then, $f$ defined by \eqref{eq:psi_lem1} becomes
\begin{equation}
f(t)=\frac{1}{(-t;q)_{\infty}}
\end{equation}
from \eqref{eq:basicexp}. Hence, we obtain
\begin{align}
\hat{B}^kf(t)&=\frac{(-1)^k}{(1-q)^k}f(t).
\end{align}

Next, we prove
\begin{equation}
\hat{D}^kc_0 =\frac{q^{k(k+\nu)}}{(q;q)_{k+\nu}}. \label{eq:Dkc0}
\end{equation}
We introduce an identity
\begin{equation}
\frac{(x;q)_\infty}{(q;q)_\nu} {}_2\phi_1\left({{0,0}\atop{q^{\nu+1}}};q,x\right)=\frac{1}{(q;q)_\nu}{}_0\phi_1\left({{-}\atop{q^{\nu+1}}};q,q^{\nu+1}x\right), \label{eq:2p1=0p1}
\end{equation}
which appears in \cite{GR}, Exercises 3.2 (iii). Note that 
\begin{align}
(q^{\nu+1};q)_{k-l}(q;q)_\nu=(q;q)_{\nu+k-l}=\frac{1}{c_{k-l}}.
\end{align}
The left hand side of \eqref{eq:2p1=0p1} is deformed as
\begin{align}
&\frac{1}{(q;q)_\nu}\left\{\sum_{k=0}^{\infty}\frac{(-1)^kq^{\frac{k(k-1)}{2}}}{(q;q)_k}x^k\right\} \left\{\sum_{k=0}^{\infty} \frac{1}{(q;q)_k(q^{\nu+1};q)_k}x^k\right\} \nonumber\\
&=\sum_{k=0}^{\infty}\left\{\sum_{l=0}^k \frac{(-1)^lq^{\frac{l(l-1)}{2}}}{(q;q)_l(q;q)_{k-l}\left((q^{\nu+1};q)_{k-l}(q;q)_\nu\right)}\right\}x^k \nonumber\\
&= \sum_{k=0}^{\infty} \frac{1}{(q;q)_k}\left\{\sum_{l=0}^k (-1)^lq^{\frac{l(l-1)}{2}}
\begin{bmatrix}
k \\ l
\end{bmatrix}
 c_{k-l} \right\} x^k \nonumber\\
&= \sum_{k=0}^{\infty} \frac{1}{(q;q)_k}\left(\hat{D}^kc_0\right)x^k.
\end{align}
The right hand side of \eqref{eq:2p1=0p1} is rewritten as
\begin{align}
\frac{1}{(q;q)_\nu}\sum_{k=0}^{\infty}\frac{q^{k(k-1)}}{(q;q)_k(q^{\nu+1};q)_k}(q^{\nu+1}x)^k &=\sum_{k=0}^{\infty}\frac{q^{k(k+\nu)}}{(q;q)_k\left((q^{\nu+1};q)_k(q;q)_\nu\right)} x^k \nonumber\\
&= \sum_{k=0}^{\infty} \frac{1}{(q;q)_k}\frac{q^{k(k+\nu)}}{(q;q)_{\nu+k}}x^k.
\end{align}
We have \eqref{eq:Dkc0} by comparing these two series.

Now, applying the $q$-Euler transformation, $\hat{J}_{\nu}^{(1)}$ is deformed as
\begin{align}
\hat{J}_{\nu}^{(1)}(t;q)=\sum_{k=0}^{\infty}\frac{1}{[k]!}\frac{q^{k(k+\nu)}}{(q;q)_{k+\nu}}t^k\frac{(-1)^k}{(1-q)^k}f(t)
=f(t)\sum_{k=0}^{\infty}\frac{(-1)^kq^{k(k+\nu)}}{(q;q)_k(q;q)_{k+\nu}}t^k, \label{eq:tempJqE}
\end{align}
which gives
\begin{equation}
J_{\nu}^{(1)}(x;q)=\left(\frac{x}{2}\right)^{\nu}\frac{1}{(-x^2/4;q)_\infty}\sum_{k=0}^{\infty}\frac{(-1)^kq^{k(k+\nu)}}{(q;q)_k(q;q)_{k+\nu}}\left(\frac{x}{2}\right)^{2k} \label{eq:JqE}
\end{equation}
by \eqref{eq:temptoJackson}. Moreover, \eqref{eq:JqE} is reduced to \eqref{eq:qEqBessel} by using \eqref{eq:JacksontoqB}. $\blacksquare$\\
(Proof of Proposition \ref{prop2.1} (ii)): 
Substituting $x=q^n$ into \eqref{eq:q-besselfnc} and \eqref{eq:qEqBessel}, we respectively obtain
\begin{align}
J_\nu(q^n)&=(1-q)^\nu q^{n\nu} \sum_{j=0}^{\infty}\frac{(-1)^j(1-q)^{2j}}{(q^2;q^2)_{j}(q^2;q^2)_{j+\nu}}q^{2nj} \label{eq:qBq^n}\\
J_\nu(q^n)&=\frac{(1-q)^\nu q^{\nu n}}{(-(1-q)^2q^{2n};q^2)_\infty}\sum_{k=0}^{\infty}\frac{(-1)^k(1-q)^{2k}}{(q^2;q^2)_k(q^2;q^2)_{k+\nu}}q^{\tilde{f}_{\nu,n}(k)}, \label{eq:qBq^nEuler}
\end{align}
where
\begin{equation}
\tilde{f}_{\nu,n}(k):=2k(k+\nu+n)=2\left(k+\frac{\nu+n}{2}\right)^2-\frac{(\nu+n)^2}{2}.
\end{equation}
Note that
\begin{align}
\frac{1}{(q^2;q^2)_m}&=1+O(q^2)\quad(q\to 0)\\
\frac{1}{(-(1-q)^2q^{2n};q^2)_\infty}&=q^{n(n-1)}\left(\frac{1}{2}+O(q)\right)\quad(q\to 0,\ n\le0).\label{eq:ev_qexp}
\end{align}

If $n\ge 1$, the term with $j=0$ in \eqref{eq:qBq^n} is dominant. Hence, we readily obtain 
\begin{equation}
J_\nu(q^n)=q^{\nu n}\left(1+O(q)\right)\quad(q\to 0).
\end{equation}

Since it is difficult to evaluate \eqref{eq:qBq^n} for $n\in\mathbb{Z}_{\le0}$, we focus on the other expression \eqref{eq:qBq^nEuler} and study the minimum value of $\tilde{f}_{\nu,n}(k)$ for $k\in\mathbb{Z}_{\ge0}$. 
If $0\ge n\ge-\nu$, we find that $\min \tilde{f}_{\nu,n}(k)=\tilde{f}_{\nu,n}(0)=0$ and that the term with $k=0$ in the summation \eqref{eq:qBq^nEuler} is dominant. Hence, we obtain 
\begin{equation}
J_\nu(q^n)=q^{\nu n+n(n-1)}\left(\frac{1}{2}+O(q)\right)\quad(q\to 0).
\end{equation}
Next, if $n<-\nu$ and $\nu+n$ is even, we find that 
\begin{equation}
\min \tilde{f}_{\nu,n}(k)=\tilde{f}_{\nu,n}\left(-\frac{\nu+n}{2}\right)=-\frac{(\nu+n)^2}{2}.
\end{equation}
Hence, the term with $k=-\frac{\nu+n}{2}$ is dominant and then
\begin{equation}
J_\nu(q^n)=(-1)^{\frac{\nu+n}{2}}q^{\frac{n(n-2)-\nu^2}{2}}\left(\frac{1}{2}+O(q)\right)\quad(q\to 0)
\end{equation}
holds. 
Finally, we study the case where $n<-\nu$ and $\nu+n$ is odd. We put $\nu+n=2\mu+1$ $(\mu=-1,-2,\dots)$. We find that
\begin{equation}
\min \tilde{f}_{\nu,n}(k)= \tilde{f}_{\nu,n}(-\mu)=\tilde{f}_{\nu,n}(-\mu-1)=-2\mu(\mu+1)=:\tilde{m}.
\end{equation}
Then, the reading term in the summation in \eqref{eq:qBq^nEuler} may be given by the sum of the terms with $k=-\mu$ and $-\mu-1$, 
\begin{align}
&(-1)^{-\mu}(1-q)^{-2\mu}q^{\tilde{m}}(1+O(q^2))+(-1)^{-\mu-1}(1-q)^{-2\mu-2}q^{\tilde{m}}(1+O(q^2))\nonumber\\
&=(-1)^{-\mu-1}q^{1+\tilde{m}}(2+O(q)).
\end{align}
We also find from $|\tilde{f}_{\nu,n}(-\mu)-\tilde{f}_{\nu,n}(-\mu+1)|=4$ that the other terms do not contribute to the reading term. Therefore, we have
\begin{align}
J_\nu(q^n)&=q^{\nu n+n(n-1)}\cdot(-1)^{-\mu-1}q^{1+\tilde{m}}\left(1+O(q)\right)\nonumber\\
&=(-1)^{\frac{\nu+n+1}{2}}q^{\frac{n(n-2)-\nu^2+3}{2}}(1+O(q))\quad(q\to 0).
\end{align}
Now, \eqref{eq:evaluation} is proved. $\blacksquare$\\
\subsection{Ultradiscrete limit and supplementary result}
If we put $x=q^n$, $q=e^{\frac{Q}{\varepsilon}}$ $(Q<0)$ and 
\begin{align}
J_\nu(q^n) = \{ s(\beta_n) - s(-\beta_n)\} e^{\frac{B_n}{\varepsilon}},
\end{align}
\eqref{eq:qBeq} is ultradiscretized as
\begin{multline}
 \max[S(\beta^\nu_{n+1})+B^\nu_{n+1},S(-\beta^\nu_n)+B^\nu_{n}-\nu Q,S(\beta^\nu_{n-1})+B^\nu_{n-1}+\max(0, (2n-2)Q)] \\
  =\max[S(-\beta^\nu_{n+1})+B^\nu_{n+1},S(\beta^\nu_n)+B^\nu_{n}-\nu Q,S(-\beta^\nu_{n-1})+B^\nu_{n-1}+\max(0, (2n-2)Q)].    \label{eq:max-bessel}
\end{multline}
Applying lemma 1 to proposition \ref{prop2.1} (ii), we readily obtain the explicit expression of the ultradiscrete analog of the $q$-Bessel function
\begin{align}
\mathcal{B}_{n}^{\nu} &= \left(\beta_{n}^{\nu}, B_{n}^{\nu}\right) \nonumber\\
&=
\begin{cases}
\left(1, n\nu Q\right)&(n\ge 1)\\
\left(1, n(n+\nu-1)Q\right)&(0\ge n\ge -\nu)\\
\left((-1)^{\frac{n+\nu}{2}}, \frac{n(n-2)-\nu^2}{2}Q\right)&(n\le -\nu-1, n+\nu:\mbox{even})\\
\left((-1)^{\frac{n+\nu+1}{2}}, \frac{n(n-2)-\nu^2+3}{2}Q\right)&(n\le -\nu-1, n+\nu:\mbox{odd}).
\end{cases}
\end{align}
We note that this limit is not identical to the ultradiscrete Bessel function by Narasaki (see Appendix A), only in the case of $n\le -\nu-1, n+\nu:$ odd. 

We note another supplementary result obtained from proposition \ref{prop2.1}. It is concerned with the number of restricted partitions $p_n(k)$ defined by the generation function
\begin{equation}
\frac{1}{(q;q)_n}=\sum_{k=0}^{\infty}p_n(k)q^k. \label{eq:GF}
\end{equation}
We have the series expression of \eqref{eq:tempJ} by utilizing \eqref{eq:GF},
\begin{align}
\hat{J}_{\nu}^{(1)}(q^n;q)&=\sum_{k=0}^{\infty}(-1)^k \left(\sum_{j=0}^{\infty}p_k(j)q^j\right)\left(\sum_{j=0}^{\infty}p_{k+\nu}(j)q^j\right)q^{kn} \nonumber \\
&=\sum_{k=0}^{\infty}\sum_{j=0}^{\infty}(-1)^k \sum_{l=0}^j p_k(j-l)p_{k+\nu}(l) q^{j+kn}. \label{eq:tempJbyp}
\end{align}
From \eqref{eq:tempJqE}, we also obtain
\begin{align}
&\hat{J}_{\nu}^{(1)}(q^n;q) =\nonumber \\
&\begin{cases}
1+O(q) & (n\ge 1)\\
q^{\frac{n(n-1)}{2}}(\frac{1}{2}+O(q)) & (0\ge n\ge-\nu)\\
(-1)^{-\frac{n+\nu}{2}}q^{\frac{n(n-1)}{2}-\left(\frac{n+\nu}{2}\right)^2}(\frac{1}{2}+O(q)) & (n<-\nu,\ n+\nu:\mbox{even}) \\
(-1)^{-\frac{n+\nu-1}{2}}q^{\frac{n(n-1)}{2}-\frac{(n+\nu-1)(n+\nu+3)}{4}}(\frac{1}{2}+O(q)) & (n<-\nu,\ n+\nu:\mbox{odd}).
\end{cases}
\end{align}
Comparing these two expressions, we find that \eqref{eq:tempJbyp} has an infinite number of extra terms for a negative integer $n$. From this fact, we have the following formula.
\begin{prop}
For given $\nu\in\mathbb{Z}_{\ge0}$ and $n\in\mathbb{Z}_{<0}$, fix $m\in\mathbb{Z}$ such that
\begin{equation}
m<
\begin{cases}
\frac{n(n-1)}{2} & (0>n\ge-\nu)\\
\frac{n(n-1)}{2}-\left(\frac{n+\nu}{2}\right)^2 & (n<-\nu,\ n+\nu:\mbox{\rm even}) \\
\frac{n(n-1)}{2}-\frac{(n+\nu-1)(n+\nu+3)}{4} & (n<-\nu,\ n+\nu:\mbox{\rm odd}).
\end{cases}
\end{equation}
Then 
\begin{equation}
\sum_{j,k} (-1)^k \sum_{l=0}^j p_k(j-l)p_{k+\nu}(l)=0
\end{equation}
holds, where $j$ and $k$ run over all pairs satisfying $j+kn=m$ in the
summation. $\blacksquare$
\end{prop}
We do not know a combinatorial meaning of this (formal) formula yet. 

\section{Special solutions of the discrete Painlev\'{e} III equation and their ultradiscrete limit}
In this section, we substitute the $q$-Bessel function into \eqref{eq:tau_det} and study the p-ultradiscrete limit of the resulting function. 
If $\tau_N^\nu(n)$ is written as the form \eqref{eq:suff_evaluation}, its p-ultradiscrete limit is readily obtained by lemma 1. We first explain useful notations and summarize the main result. 

We rewrite \eqref{eq:evaluation} into a simpler form since we often use it. We introduce 
\begin{align}
\psi(n)&:=n(n+\nu-1) \label{eq:psi}\\
p_1(n) &:= \frac{3\{1+(-1)^{n+1}\}}{4} = 
\begin{cases}
0 & (n\text{: even}) \\
\frac{3}{2} & (n\text{: odd})
\end{cases} \label{eq:p_1}
\\
p_2(n) &:= \frac{3+(-1)^{n+1}}{4} = 
\begin{cases}
\frac{1}{2} & (n\text{: even}) \\
1 & (n\text{: odd})
\end{cases}\label{eq:p_2} \\
\varphi_\nu(n) &:=\frac{n(n-2)}{2}-\frac{\nu^2}{2} \label{eq:phi_nu} 
\end{align}
and the binomial coefficient $\binom{n}{2}=(n-1)n/2$. By using \eqref{eq:rmJ} and these notations, \eqref{eq:evaluation} is rewritten as 
\begin{align}
\mathrm{J}_\nu(n)=
{}&q^{n\nu}\left(1+O(q)\right)&&(n\ge 1) \label{eq:ev_n>>1}\\
\mathrm{J}_\nu(n)={}&q^{\psi (n)}\left(\frac{1}{2}+O(q)\right)&&(0\ge n\ge -\nu) \label{eq:ev_n-midd}\\
\mathrm{J}_\nu(n)= {}&(-1)^{\binom{n+\nu+1}{2}}q^{\varphi_\nu(n)+p_1(n)}\left(p_2(n)+O(q)\right) && (n\le -\nu-1). \label{eq:ev_n<<-1}
\end{align}

For $\tau_N^\nu(n)$, we define the sign variable $y_{N,n}^\nu$ and amplitude variable $Y_{N,n}^\nu$ by
\begin{align}
y_{N,n}^\nu = \frac{\tau_N^\nu(n)}{|\tau_N^\nu(n)|},\quad
 |\tau_N^\nu(n)| = e^{\frac{Y_{N,n}^\nu}{\varepsilon}}, 
\end{align}
respectively. Then, p-ultradiscrete analogs of \eqref{eq:bilinear1}--\eqref{eq:bilinear4} are respectively written as
\begin{align}\label{eq:UDbilinear1}
\max[&S(y_{N+1}^{\nu,n}y_N^{\nu+1,n+1})+Y_{N+1}^{\nu,n}+Y_N^{\nu+1,n+1},\nonumber\\
&S(-y_{N+1}^{\nu,n+1}y_N^{\nu+1,n})+Y_{N+1}^{\nu,n+1}+Y_N^{\nu+1,n}-(\nu+N)Q,\nonumber\\
&S(y_{N+1}^{\nu+1,n}y_N^{\nu,n+1})+Y_{N+1}^{\nu+1,n}+Y_N^{\nu,n+1}+(n+2N)Q,\nonumber\\
&S(-y_{N+1}^{\nu+1,n}y_N^{\nu,n+1})+Y_{N+1}^{\nu+1,n}+Y_N^{\nu,n+1}+(n+2N+1)Q]\nonumber\\
=\max[&S(-y_{N+1}^{\nu,n}y_N^{\nu+1,n+1})+Y_{N+1}^{\nu,n}+Y_N^{\nu+1,n+1},\nonumber\\
&S(y_{N+1}^{\nu,n+1}y_N^{\nu+1,n})+Y_{N+1}^{\nu,n+1}+Y_N^{\nu+1,n}-(\nu+N)Q,\nonumber\\
&S(-y_{N+1}^{\nu+1,n}y_N^{\nu,n+1})+Y_{N+1}^{\nu+1,n}+Y_N^{\nu,n+1}+(n+2N)Q,\nonumber\\
&S(y_{N+1}^{\nu+1,n}y_N^{\nu,n+1})+Y_{N+1}^{\nu+1,n}+Y_N^{\nu,n+1}+(n+2N+1)Q],
\end{align}
\begin{align}\label{eq:UDbilinear2}
\max[&S(y_{N+1}^{\nu+1,n}y_N^{\nu,n+1})+Y_{N+1}^{\nu+1,n}+Y_N^{\nu,n+1},\nonumber\\
&S(-y_{N+1}^{\nu+1,n+1}y_N^{\nu,n})+Y_{N+1}^{\nu+1,n+1}+Y_N^{\nu,n}+(\nu-N+1)Q,\nonumber\\
&S(-y_{N+1}^{\nu,n}y_N^{\nu+1,n+1})+Y_{N+1}^{\nu,n}+Y_N^{\nu+1,n+1}+(n+2N)Q,\nonumber\\
&S(y_{N+1}^{\nu,n}y_N^{\nu+1,n+1})+Y_{N+1}^{\nu,n}+Y_N^{\nu+1,n+1}+(n+2N+1)Q]\nonumber\\
=\max[&S(-y_{N+1}^{\nu+1,n}y_N^{\nu,n+1})+Y_{N+1}^{\nu+1,n}+Y_N^{\nu,n+1},\nonumber\\
&S(y_{N+1}^{\nu+1,n+1}y_N^{\nu,n})+Y_{N+1}^{\nu+1,n+1}+Y_N^{\nu,n}+(\nu-N+1)Q,\nonumber\\
&S(y_{N+1}^{\nu,n}y_N^{\nu+1,n+1})+Y_{N+1}^{\nu,n}+Y_N^{\nu+1,n+1}+(n+2N)Q,\nonumber\\
&S(-y_{N+1}^{\nu,n}y_N^{\nu+1,n+1})+Y_{N+1}^{\nu,n}+Y_N^{\nu+1,n+1}+(n+2N+1)Q],
\end{align}
\begin{align}\label{eq:UDbilinear3}
\max[&S(y_{N+1}^{\nu,n}y_N^{\nu+1,n+3})+Y_{N+1}^{\nu,n}+Y_N^{\nu+1,n+3},\nonumber\\
&S(-y_{N+1}^{\nu,n+1}y_N^{\nu+1,n+2})+Y_{N+1}^{\nu,n+1}+Y_N^{\nu+1,n+2}-(\nu+N)Q,\nonumber\\
&S(y_{N+1}^{\nu+1,n}y_N^{\nu,n+3})+Y_{N+1}^{\nu+1,n}+Y_N^{\nu,n+3}+nQ,\nonumber\\
&S(-y_{N+1}^{\nu+1,n}y_N^{\nu,n+3})+Y_{N+1}^{\nu+1,n}+Y_N^{\nu,n+3}+(n+1)Q]\nonumber\\
=\max[&S(-y_{N+1}^{\nu,n}y_N^{\nu+1,n+3})+Y_{N+1}^{\nu,n}+Y_N^{\nu+1,n+3},\nonumber\\
&S(y_{N+1}^{\nu,n+1}y_N^{\nu+1,n+2})+Y_{N+1}^{\nu,n+1}+Y_N^{\nu+1,n+2}-(\nu+N)Q,\nonumber\\
&S(-y_{N+1}^{\nu+1,n}y_N^{\nu,n+3})+Y_{N+1}^{\nu+1,n}+Y_N^{\nu,n+3}+nQ,\nonumber\\
&S(y_{N+1}^{\nu+1,n}y_N^{\nu,n+3})+Y_{N+1}^{\nu+1,n}+Y_N^{\nu,n+3}+(n+1)Q],
\end{align}
\begin{align}\label{eq:UDbilinear4}
\max[&S(y_{N+1}^{\nu+1,n}y_N^{\nu,n+3})+Y_{N+1}^{\nu+1,n}+Y_N^{\nu,n+3},\nonumber\\
&S(-y_{N+1}^{\nu+1,n+1}y_N^{\nu,n+2})+Y_{N+1}^{\nu+1,n+1}+Y_N^{\nu,n+2}+(\nu-N+1)Q,\nonumber\\
&S(-y_{N+1}^{\nu,n}y_N^{\nu+1,n+3})+Y_{N+1}^{\nu,n}+Y_N^{\nu+1,n+3}+nQ,\nonumber\\
&S(y_{N+1}^{\nu,n}y_N^{\nu+1,n+3})+Y_{N+1}^{\nu,n}+Y_N^{\nu+1,n+3}+(n+1)Q]\nonumber\\
=\max[&S(-y_{N+1}^{\nu+1,n}y_N^{\nu,n+3})+Y_{N+1}^{\nu+1,n}+Y_N^{\nu,n+3},\nonumber\\
&S(y_{N+1}^{\nu+1,n+1}y_N^{\nu,n+2})+Y_{N+1}^{\nu+1,n+1}+Y_N^{\nu,n+2}+(\nu-N+1)Q,\nonumber\\
&S(y_{N+1}^{\nu,n}y_N^{\nu+1,n+3})+Y_{N+1}^{\nu,n}+Y_N^{\nu+1,n+3}+nQ,\nonumber\\
&S(-y_{N+1}^{\nu,n}y_N^{\nu+1,n+3})+Y_{N+1}^{\nu,n}+Y_N^{\nu+1,n+3}+(n+1)Q].
\end{align}
These ultradiscrete equations are solved by the p-ultradiscrete limit of \eqref{eq:tau_det} presented below. In order to give the explicit functional form, we introduce the following six cases: 
\begin{itemize}
\item[(A)] $n\ge 1$,
\item[(B)] $0\ge n\ge \max(2-2N, -\nu-N)+1$,
\item[(C-a)] $\nu \ge N-1$ and $2-2N  \ge n \ge 1-\nu-N $,
\item[(C-b)] $\nu \le N-2$ and $-\nu-N \ge n\ge 3-2N$,
\item[(D)] $\min(2-2N, -\nu-N)\ge n \ge 2-2N-\nu $,
\item[(E)] $n\le 1-2N-\nu$.
\end{itemize}
Note that there exist $n$'s such that all anti-diagonal elements in \eqref{eq:tau_det} are type of \eqref{eq:ev_n-midd} for $\nu \ge N-1$ but do not for $\nu \le N-2$. We also introduce
\begin{align}
A_{N,n} &:= 2n+3N-3 \\
B_\nu(k) &:= \varphi_\nu(k) +p_1(k+\nu) \\
M &=\min(\mathrm{floor}(|n|/2)+1,N), \label{eq:def_M}
\end{align}
where $\mathrm{floor}(x)$ denotes the integer part of $x$. Now, the p-ultradiscrete limit of \eqref{eq:tau_det} is written as follows. \\
Case (A):
\begin{align}
y_{N,n}^\nu &= (-1)^{\binom{N}{2}}\\
Y_{N,n}^\nu &= Q\left\{ \frac{\nu N}{2} A_{N,n}+2\binom{N}{2}(n+N-2) \right\}.
\end{align}
Case (B):
\begin{align}
y_{N,n}^\nu &= (-1)^{\binom{N}{2}}\\
Y_{N,n}^\nu &= Q\biggl\{ \frac{\nu N}{2} A_{N,n}+2 \binom{N-M}{2}(n+N+M-2) \nonumber\\
&\phantom{=}+\frac{1}{2}\binom{M+1}{3}+2M\binom{n+N+\frac{M-3}{2}}{2} \biggr\}.
\end{align}
Case (C-a):
\begin{align}
y_{N,n}^\nu &= (-1)^{\binom{N}{2}}\\
Y_{N,n}^\nu &= Q\left\{ \frac{N}{4}A_{N,n}(A_{N,n}+2\nu-2)+\frac{1}{2}\binom{N+1}{3} \right\}.
\end{align}
Case (C-b):
\begin{align}
y_{N,n}^\nu &= (-1)^{\binom{N}{2}} \prod_{k=n+N-1}^{-\nu-1} (-1)^{\binom{k+\nu+1}{2}}\\
Y_{N,n}^\nu &= Q\biggl\{ \frac{\nu N}{2} A_{N,n}+2 \binom{N-M}{2}(n+N+M-2) \nonumber \\
 &\phantom{=} + \sum_{k=-\nu}^{n+N+M-2} k(k-1) + \sum_{k=n+N-1}^{-\nu-1} \left(B_\nu(k)-\nu k\right) \biggr\}.
\end{align}
Case (D):
\begin{align}
y_{N,n}^\nu &= (-1)^{\binom{N}{2}} \prod_{k=n+N-1}^{-\nu-1} (-1)^{\binom{k+\nu+1}{2}}\\
Y_{N,n}^\nu &= Q\left\{ \sum_{k=n+N-1}^{-\nu-1} B_\nu(k) + \sum_{k=-\nu}^{n+2N-2} \psi(k)  \right\}.
\end{align}
Case (E):
\begin{align}
y_{N,n}^\nu &= (-1)^{\binom{N}{2}} \prod_{k=n+N-1}^{n+2N-2} (-1)^{\binom{k+\nu+1}{2}}\\
Y_{N,n}^\nu &= Q \sum_{k=n+N-1}^{n+2N-2} B_\nu(k).
\end{align}
We shall study each case in the following subsections. 
\subsection{Case (A)}
In this case, all arguments of $\mathrm{J}_\nu(n)$ in \eqref{eq:tau_det} are positive. The technique developed in \cite{IST} is useful in fact. Substituting \eqref{eq:qBq^n} into \eqref{eq:tau_det} and using multi-linearity of determinant, we obtain 
\begin{align}
\tau_N^\nu(n) = &(1-q)^{\nu N} \left\{\prod_{k=1}^{N} q^{(n+3k-3)\nu}\right\} \sum_{j_1,\ldots, j_N} P_N^\nu(\mbox{\boldmath $j$}) \left\{\prod_{k=1}^{N} q^{2(n+2k-2)j_k}\right\}\nonumber \\
&\times \begin{vmatrix}
1 & q^{2j_1} & \dots & q^{2(N-1)j_1} \\
1 & q^{2j_2} & \dots & q^{2(N-1)j_2} \\
\vdots & \vdots & \ddots & \vdots \\
1 & q^{2j_N} & \dots & q^{2(N-1)j_N}  \\
\end{vmatrix}, \label{eq:tau_det_qB}
\end{align}
where
\begin{align}
P_N^\nu(\boldsymbol{j}) &:= \prod_{k=1}^N \frac{(-1)^{j_k} (1-q)^{2j_k}}{(q^2;q^2)_{j_k+\nu}(q^2;q^2)_{j_k}} = \left\{\prod_{k=1}^N (-1)^{j_k}\right\} (1+O(q))\label{eq:P}\\
\sum_{j_1,\ldots, j_N} &:= \sum_{j_1=0}^\infty\sum_{j_2=0}^\infty\dots\sum_{j_N=0}^\infty. \label{eq:sums}
\end{align}
Since the determinant in \eqref{eq:tau_det_qB} is the Vandermonde determinant, the terms with $j_k = j_l$ $(k \neq l)$ disappear. Moreover, under any permutation of $(j_1, j_2, \dots, j_N)$, the absolute values of all elements but the factor $\prod_{k=1}^{N} q^{2(n+2k-2)j_k}$ in \eqref{eq:tau_det_qB} are invariant. Therefore, noting $0<q<1$, the largest absolute value of the monomial in \eqref{eq:tau_det_qB} is achieved by the $j_k$'s with
\begin{align}
0\le j_N < \dots < j_2 < j_1, \label{eq:j_ineq.}
\end{align}
which maximize $\prod_{k=1}^{N} q^{2(n+2k-2)j_k}$. Then, in the Vandermonde determinant, the product of diagonal elements $\prod_{k=1}^{N} q^{2(k-1)j_k}$ contributes to the evaluation. Hence, we may consider $j_k$'s which maximize $\prod_{k=1}^{N} q^{2(n+2k-2)j_k + 2(k-1)j_k}$, that is, minimize
\begin{align}
\sum_{k=1}^{N} 2(n+3k-3)j_k
\end{align}
under \eqref{eq:j_ineq.}. We readily find that such $j_k$'s are given by
\begin{align}
j_k=N-k,\quad k=1,2,\dots,N.
\end{align}
Moreover, further contribution of the factor $\prod_{k=1}^{N} q^{(n+3k-3)\nu}$ and sign $\prod_{i=1}^N (-1)^{N-i} = (-1)^{\binom{N}{2}}$ from $P_N^\nu(\boldsymbol{j})$ must be considered. Accordingly, the reading term of $\tau_N^\nu(n)$ is given by
\begin{align}
\tau_N^\nu(n) &\sim (-1)^{\binom{N}{2}} \prod_{k=1}^{N} q^{(n+3k-3)(2N-2k+\nu)}\nonumber\\
&=(-1)^{\binom{N}{2}} q^{\frac{\nu N}{2} (2n+3N-3)+N(N-1)(n+N-2)}.
\end{align}
\subsection{Case (C-a), (D) and (E)} \label{ssec:C-a}
In these cases, all arguments of $\mathrm{J}_\nu(n)$ in anti-diagonal elements of \eqref{eq:tau_det} are negative. The following proposition plays an quite important role for evaluation.
\begin{prop}\label{prop:2}
 Write the absolute value of the reading term of $\mathrm{J}_\nu(n)$ as $q^{g_\nu(n)}$. For $n \in \mathbb{Z}$ and $k,l \in \mathbb{Z}_{>0}$, 
\begin{align}
\tilde{g} := g_\nu(n) + g_\nu(n+2k+l) - g_\nu(n+2k) - g_\nu(n+l) 
\begin{cases}
=0 & (n\ge 0) \\
>0 & (n\le -1)
\end{cases}
\end{align}
holds. \\
\end{prop}
(proof) This inequality is proved by direct calculation with considering all possible 46 cases of $g_\nu$'s. We here illustrate typical three cases. Firstly, if $n+2k+l \le -\nu-1$ and both of $n+\nu$ and $n+l+\nu$ are even, we have
\begin{align*}
g_\nu (n) &= \varphi_\nu (n), &
g_\nu (n+2k+l) &= \varphi_\nu (n+2k+l), \nonumber \\ 
g_\nu (n+l) &= \varphi_\nu (n+l), &
g_\nu (n+2k) &= \varphi_\nu (n+2k)
\end{align*}
and $\tilde{g} = 2kl \ge 2 >0$. Secondly, if $n+2k\le -\nu-1$, $n$ is even, $0\ge n+l \ge -\nu$ and $n+2k+l\ge 1$, then we have
\begin{align*}
g_\nu (n) &= \varphi_\nu(n), &
g_\nu (n+2k+l) &= \nu(n+2k+l), \nonumber \\
g_\nu (n+l) &= \psi (n+l), &
g_\nu (n+2k) &= \varphi_\nu (n+2k).
\end{align*}
Since $n+l+\nu-1 \ge -1$, we consider two cases, $n+l+\nu-1 \ge 0$ and $n+l+\nu = 0$. When $n+l+\nu-1 \ge 0$, $g_\nu (n+l)\le 0$ and therefore $\tilde{g}\ge g_\nu (n) + g_\nu (n+2k+l) - g_\nu (n+2k) = -2k(n+k)+\nu(n+2k+l)+2k >0$. When $n+l+\nu = 0$, we obtain $2k-\nu\ge 1$ and therefore $\tilde{g}= -2k(n+k) +\nu(2k-\nu)+2k >0$. Finally, if $n\le -\nu-1$ and $n+\nu$ is even and $0\ge n+2k, n+l, n+2k+l \ge -\nu$, then $g_\nu$'s are given by
\begin{align*}
g_\nu (n) &= \varphi_\nu (n), &
g_\nu (n+2k+l) &= \psi (n+2k+l), \nonumber \\
g_\nu (n+l) &= \psi (n+l), &
g_\nu (n+2k) &= \psi (n+2k)
\end{align*}
and $\tilde{g}= \{8kl-(n+\nu)^2\}/2$. Noting that $|n+\nu|\le l$ and $|n+\nu|\le 2k$, we obtain $(n+\nu)^2\le 2kl <8kl$. Hence, $\tilde{g}>0$ holds. $\blacksquare$

From proposition 3, the inequality $|\mathrm{J}_\nu(n) \mathrm{J}_\nu(n+2k+l)| < |\mathrm{J}_\nu(n+2k) \mathrm{J}_\nu(n+l)|$ for $n<0$ as $q\to 0$ follows. 
This inequality implies that the product of anti-diagonal elements is dominant in \eqref{eq:tau_det} for $-n \gg 1$ (See \eqref{eq:anti-diag.}).
\begin{align}
\begin{vmatrix}
\\
\cdots & \mathrm{J}_\nu(n) & \cdots & \mathrm{J}_\nu(n+l) & \cdots \\
\vdots & \vdots & \ddots & \vdots & \vdots \\
\cdots & \mathrm{J}_\nu(n+2k) & \cdots & \mathrm{J}_\nu(n+2k+l) & \cdots
\\
\\
\end{vmatrix}.\label{eq:anti-diag.}
\end{align}
Moreover, this is true even for the case where all anti-diagonal elements have non-positive arguments.

We first study the case (E). In this case, all anti-diagonal elements of \eqref{eq:tau_det} are type of \eqref{eq:ev_n<<-1} and therefore we have
\begin{align}
\tau_N^\nu(n) \sim (-1)^{\binom{N}{2}} \prod_{k=n+N-1}^{n+2N-2} (-1)^{\binom{k+\nu+1}{2}} p_2(k+\nu) q^{\varphi_\nu(k) +p_1(k+\nu)},
\end{align}
where we use notation \eqref{eq:p_1}--\eqref{eq:phi_nu}.

For the case (C-a), all anti-diagonal elements in \eqref{eq:tau_det} are type of \eqref{eq:ev_n-midd}. The reading term of \eqref{eq:tau_det} is given by 
\begin{align}
\tau_N^\nu(n) &\sim (-1)^{\binom{N}{2}} \prod_{k=n+N-1}^{n+2N-2} \frac{1}{2} q^{\psi(k)} \nonumber \\
&= (-1)^{\binom{N}{2}} 2^{-N} q^{\frac{\nu N}{2} (2n+3N-3) + \frac{N}{4}(2n+3N-3)(2n+3N-5)+\frac{(N-1)N(N+1)}{12}}\nonumber \\
&= (-1)^{\binom{N}{2}} 2^{-N} q^{\frac{N}{4}(2n+3N-3)(2n+3N+2\nu-5)+\frac{1}{2}\binom{N+1}{3}}, \label{eq:result_Ca}
\end{align}
where $\psi$ is defined by \eqref{eq:psi}.

The case (D) gives a \textit{mixed} situation. The anti-diagonal elements in \eqref{eq:tau_det} are type of \eqref{eq:ev_n-midd} or \eqref{eq:ev_n<<-1}. The reading term is, by using \eqref{eq:psi}--\eqref{eq:phi_nu}, represented as
\begin{align}
\tau_N^\nu(n)&\sim (-1)^{\binom{N}{2}} \prod_{k=n+N-1}^{-\nu-1} (-1)^{\binom{k+\nu+1}{2}} p_2(k+\nu) q^{\varphi_\nu(k) +p_1(k+\nu)} \prod_{k=-\nu}^{n+2N-2} \frac{1}{2}q^{\psi(k)}. \label{eq:D_result}
\end{align}
\subsection{Case (B)} \label{sec:caseB}
In this case, $\mathrm{J}_\nu(n)$'s of type \eqref{eq:ev_n>>1} and \eqref{eq:ev_n-midd} appear as elements in \eqref{eq:tau_det}. Moreover, there exist some rows (or exists a row) which consist only of $\mathrm{J}_\nu(n)$'s with positive arguments. The product of anti-diagonal elements is not dominant any more. 
We consider $M$ defined by \eqref{eq:def_M}. Then, the $i$th row $(M+1\le i\le N)$ consists only of $\mathrm{J}_\nu(n)$'s with positive arguments. 
When $n$ is even, \eqref{eq:tau_det} has the form
\begin{align}
\begin{array}{r}
\\
\\
(M\mbox{th})\\
\\
\\
\ 
\end{array}
\left|
\begin{array}{cccc}
\mathrm{J}_\nu(n) & \mathrm{J}_\nu(n+1) & \cdots & \mathrm{J}_\nu(n+N-1) \\
\vdots & \vdots &  & \vdots \\
\mathrm{J}_\nu(0) & \mathrm{J}_\nu(1) & \cdots & \mathrm{J}_\nu(N-1) \\
\mathrm{J}_\nu(2) & \mathrm{J}_\nu(3) & \cdots & \mathrm{J}_\nu(N+1) \\
\vdots & \vdots & \ddots & \vdots \\
\mathrm{J}_\nu(n+2N-2)& \mathrm{J}_\nu(n+2N-1) & \cdots & \mathrm{J}_\nu(n+3N-3) 
\end{array}
\right|.
\end{align}
We shall present the procedure to calculate its reading term. Although we illustrate for the case where some anti-diagonal elements are type of \eqref{eq:ev_n-midd}, this procedure is available for the case where all anti-diagonal elements are type of \eqref {eq:ev_n>>1} or for odd $n$ in a similar manner. We use notation $\{n\}:=(-(1-q)^2q^{2n};q^2)_\infty$ for simplicity. Substituting the series expressions of $\mathrm{J}_\nu(q^n)$ and employing multi-linearity of determinant, we obtain 
\begin{align}
\tau_N^\nu(n) = (1-q)^{\nu N} \left\{\prod_{k=1}^{N} q^{(n+3k-3)\nu} \right\}
\sum_{j_1,\ldots, j_N} 
P_N^\nu(\boldsymbol{j}) \left\{\prod_{k=M+1}^{N} q^{ 2(n+2k-2)j_k}\right\} \nonumber \\
\times
\begin{vmatrix}
\frac{q^{2j_1(j_1+\nu+n)}}{\{n\}} & \frac{q^{2j_1(j_1+\nu+n+1)}}{\{n+1\}} & \cdots & \frac{q^{2j_1(j_1+\nu+n+N-1)}}{\{n+N-1\}}\\
\vdots & \vdots & \frac{q^{2j_k(j_k+\nu+n+2\kappa+\lambda-3)}}{\{n+2\kappa+\lambda-3\}} & \vdots \\
\frac{q^{2j_M(j_M+\nu)}}{\{0\}} & q^{2j_M} & \cdots & q^{2(N-1)j_M}\\
1 & q^{2j_{M+1}} & \dots & q^{2(N-1)j_{M+1}} \\
\vdots & \vdots & \ddots & \vdots \\
1 & q^{2j_N} & \dots & q^{2(N-1)j_N}  \\
\end{vmatrix}, \label{eq:tau_det_caseB}
\end{align}
where we use \eqref{eq:P} and \eqref{eq:sums}. 
Let the $(\kappa, \lambda)$-element for the determinant in \eqref{eq:tau_det_caseB} be of type \eqref{eq:ev_n-midd}. It has the form $q^{2j_\kappa(j_\kappa+\nu+n+2\kappa+\lambda-3)}/\{n+2\kappa+\lambda-3\}$ and $\nu+n+2\kappa+\lambda-3\ge 0$ (see section 2). Hence, the minimum of $j(j+\nu+n+2\kappa+\lambda-3)$ is achieved at $j=0$. Note that the determinants with $j_k=j_l$ $(1\le k, l \le M)$ do not disappear differently from the case (A). Therefore, the largest absolute value of the monomial in this determinant may be achieved by the $j_k$'s with
\begin{align}
j_k=
\begin{cases}
0 & (1\le k\le M, k=N)\\
N-k & (M+1\le k \le N-1).
\end{cases}\label{eq:setofjs}
\end{align}
Although other sets of $j_k$'s actually achieve the largest absolute value, we discuss them later. 
Noting that $(1-q)^{k} \sim 1$ and $P_N^\nu(\boldsymbol{j}) \sim (-1)^{\sum_{k=M+1}^{N-1}(N-k)} = (-1)^{\binom{N-M}{2}}$ under \eqref{eq:setofjs}, our aim turns into evaluating 
\begin{align}
\tau_N^\nu(n) \sim (-1)^{\binom{N-M}{2}} \left\{\prod_{k=1}^{N} q^{(n+3k-3)\nu} \right\} \left\{\prod_{k=M+1}^{N} q^{2(n+2k-2)(N-k)} \right\} \tilde{\tau}_N^\nu(n), 
\end{align}
where
\begin{align}
\tilde{\tau}_N^\nu(n) &:=
\begin{vmatrix}
\langle n\rangle & \langle n+1\rangle & \cdots & \langle n+N-1\rangle\\
\vdots & \vdots & \langle n+2\kappa+\lambda-3\rangle & \vdots \\
\langle 0\rangle & 1 & \cdots & 1\\
1 & q^{2(N-M-1)} & \dots & q^{2(N-1)(N-M-1)} \\
1 & q^{2(N-M-2)} & \dots & q^{2(N-1)(N-M-2)} \\
\vdots & \vdots & \ddots & \vdots \\
1 & 1 & \dots & 1  \\
\end{vmatrix} \\
\langle n \rangle &:= 1/\{n\} \quad (n\le 0). \label{eq:<n>}
\end{align}
We deform $\tilde{\tau}_N^\nu(n)$ according to \textit{Procedure} presented in Appendix B. 
In order to represent the resulting expression, we introduce some notations. For $x_k=q^{2(N-M-k)}$, $k=1,2,\ldots, N-M-1$, we consider the fundamental symmetric expression $\tilde{a}_k$ among them, that is,
\begin{align*}
&\tilde{a}_0\equiv 1, \quad
\tilde{a}_1 = -(x_1+\cdots +x_{N-M-1}), \quad
\tilde{a}_2 = x_1x_2+\cdots +x_{N-M-2}x_{N-M-1}, \\
& \dots,\quad \tilde{a}_k = (-1)^k \sum x_{k_1}\cdots x_{k_k},\quad \dots ,\\
&\tilde{a}_{N-M-1} = (-1)^{N-M-1} x_1x_2\cdots x_{N-M-1}, \quad
\tilde{a}_{N-M} \equiv 0.
\end{align*}
Note that 
\begin{align}
\tilde{a}_k\sim (-1)^k q^{2+4+\cdots + 2k} = (-1)^k q^{k(k+1)} \label{eq:a_k-eval}
\end{align}
as $q\to 0$. We further add a new variable $x_{N-M}=1$ to $x_1,\ldots, x_{N-M-1}$ and write the fundamental symmetric expression among them as $a_k$ $(k=1,2,\ldots, N-M)$ and set $a_0=1$. Then, the relations
\begin{align}
a_k= \tilde{a}_k - \tilde{a}_{k-1} \quad (k=1,2,\ldots, N-M)\label{eq:symm_relation}
 \end{align} 
hold. For convenience, we extend definition of $\langle n \rangle$ defined in \eqref{eq:<n>} as
\begin{align}
\langle n \rangle = 
\begin{cases}
1/\{n\} & (n\le 0) \\
1 & (n\ge 1).
\end{cases}
\end{align}
Using these notations, we define a function
\begin{align}
T(n) = \sum_{k=0}^{N-M} a_k (-1+\langle n-k \rangle).
\end{align}
Then, $\tilde{\tau}_N^\nu(n)$ after applying \textit{Procedure} is written as 
\begin{align}
\tilde{\tau}_N^\nu(n) = (-1)^{M(N-M)} \left\{\prod_{k=1}^{N-M-1} q^{2k(N-M-1-k)}\right\} \left\{\prod_{k=1}^{N-M-1}(q^2;q^2)_k\right\} \nonumber \\
\times \det\left(T(n+2\kappa+\lambda-3) \right)_{{1\le \kappa\le M \atop N-M+1\le \lambda \le N}}, \label{eq:tilde_tau_MM}
\end{align}
where we use trivial formula $(-1)^{2(N-M)}=1$ for simplicity. 
Since $(q^2;q^2)_k \sim 1$, our aim is boiled down to calculate the reading term of $\det\left(T(n+2\kappa+\lambda-3)\right)$. 
The following lemma tells us that the reading term of $\det\left(
T(n+2\kappa+\lambda-3)\right)$ is given by the product of anti-diagonal elements.
\begin{lem}
(i) $T(n)$ is evaluated as
\begin{align}
T(n) \sim
\begin{cases}
\frac{1}{2} q^{n(n-1)} & (n\le 0) \\
\frac{(-1)^{n-1}}{2} q^{n(n-1)} & (n\ge 1).
\end{cases}
\end{align}
(ii) Denote $\psi_0(n) = n(n-1)$. For $n\in \mathbb{Z}$ and $k,l \in \mathbb{Z}_{>0}$, \begin{align}
\psi_0(n) + \psi_0(n+k+l)-\psi_0(n+k)-\psi_0(n+l)<0 \label{eq:psi0_ineq}
\end{align}
holds. 
\end{lem}
(proof) We firstly prove (i). For $n\le 0$, we have
\begin{align}
T(n) = -\sum_{k=0}^{N-M} a_k + \sum_{k=0}^{N-M} a_k \langle n-k \rangle. 
\end{align}
The first summation becomes zero by \eqref{eq:symm_relation} and we find from \eqref{eq:ev_qexp} that $T\sim a_0 \langle n \rangle \sim \frac{1}{2} q^{n(n-1)}$. For $n\ge 1$, employing \eqref{eq:symm_relation}, \eqref{eq:a_k-eval} and \eqref{eq:ev_qexp}, we obtain 
\begin{align}
T(n) &= a_n (-1+\langle 0 \rangle) + a_{n+1} (-1+\langle -1 \rangle) + \cdots \nonumber \\
&= (\tilde{a}_n - \tilde{a}_{n-1}) \left(-\frac{1}{2}+o(1)\right)+\cdots \nonumber \\
&\sim  \frac{1}{2} \tilde{a}_{n-1} \sim \frac{(-1)^{n-1}}{2} q^{n(n-1)}. \label{eq:T_eval}
\end{align}
(ii) is proved by direct calculation. $\blacksquare$

Therefore, the reading term of $\det\left(T(n+2\kappa+\lambda-3) \right)$ is written as
\begin{align}
(-1)^{\binom{M}{2}} \prod_{k=n+N-1}^{\min (0,n+N+M-2)} \frac{1}{2}q^{k(k-1)} \prod_{k=\max (1, n+N-1)}^{n+N+M-2} (-1)^{k-1} \frac{1}{2}q^{k(k-1)}
\end{align}
and then, that of $\tilde{\tau}_n^\nu(n)$ is obtained from \eqref{eq:tilde_tau_MM}. However, when $n+N+M-2 \ge 1$, that is, $T(n)$ with $n\ge 2$ appears in the anti-diagonal elements of $\det\left( T(n+2\kappa+\lambda-3) \right)$, the reading term of $\tau_n^\nu(n)$ is not obtained only from $\tilde{\tau}_n^\nu(n)$. We must consider other determinants which have the same order of reading term, in other words, other sets of $j_k$'s. We give the following lemma:

\begin{lem}
Suppose that an anti-diagonal element in $\det\left(T(n+2\kappa+\lambda-3) \right)$ is $T(\tilde{n})$ with $\tilde{n} \ge 2$ and it is in the $i$-th row ($1\le i\le M$). In \eqref{eq:tau_det_caseB}, consider the determinants given by the sets of arguments
\begin{align}
j_k&=
\begin{cases}
1, 2, \ldots, \tilde{n}-1 & (k=i) \\
0 & (1\le k\le i,\ i+1\le k \le M)\\
N-k & (M+1\le k \le N). \\
\end{cases} \label{eq:j_ks-2}
\end{align}
The order of reading terms of these determinants is identical to that of obtained from \eqref{eq:setofjs}. 
\end{lem}
(proof) Since we find by observation that other sets of arguments give a larger order of reading term than that of \eqref{eq:setofjs}, the cases which we should study are \eqref{eq:j_ks-2} only. We write $j_i=m$ $(1\le m \le \tilde{n}-1)$ and adopt \eqref{eq:j_ks-2} for \eqref{eq:tau_det_caseB}, and then apply \textit{Procedure} but replace `step 0-$i$' with `Add the $(N-m)$th row $\times$ $(-q^{-2m(2-\tilde{n}-2i)})$'. Then we obtain a similar expression as \eqref{eq:tilde_tau_MM} but $T$'s in the $i$-th row are replaced with $T^{(m)}$ defined by
\begin{align}
T^{(m)} (n) = \sum_{k=n}^{N-M} a_k (-q^{2m(n-k)}+\langle n-k \rangle q^{2m(m+\nu+n-k)}).
\end{align}
Since $q^{2m(n-k)}\gg \langle n-k \rangle q^{2m(m+\nu+n-k)}$, our interest is to evaluate $\sum -a_k q^{2m(n-k)}$. For $\tilde{n}\ge 1$, we find from \eqref{eq:symm_relation} and \eqref{eq:a_k-eval} that
\begin{align}
a_k(-q^{2m(\tilde{n}-k)}) 
\sim \tilde{a}_{k-1}q^{2m(\tilde{n}-k)} \sim (-1)^{k-1}q^{k(k-1)+2m(\tilde{n}-k)}.
\end{align}
We put $h(k)=k(k-1)+2m(\tilde{n}-k)$ and consider $k=\tilde{n}+1, \tilde{n}+2, \ldots, N-M$. From $m\le \tilde{n}-1$ and $k\le \tilde{n}+1$, we readily obtain $m-k+1<0$ and therefore $h(k-1)-h(k)=2(m-k+1)<0$. This inequality means that 
\begin{align}
T^{(m)}(\tilde{n}) \sim \tilde{a}_{\tilde{n}-1} \sim (-1)^{\tilde{n}-1}q^{\tilde{n}(\tilde{n}-1)} \sim 2 T(\tilde{n}), 
\end{align}
which is independent of $m$ and moreover of $i$. $\blacksquare$

If we change the original set of $j_k$'s \eqref{eq:setofjs} into one of \eqref{eq:j_ks-2}, the extra sign $(-1)^{j_i}$ arises from $P_N^\nu(\boldsymbol{j})$. Hence, for $\tilde{n}\ge 2$, we may replace \eqref{eq:T_eval} with 
\begin{align}
T(\tilde{n}) \sim (-1)^{\tilde{n}-1}\frac{1}{2} q^{\tilde{n}(\tilde{n}-1)} + (-1)^{\tilde{n}-1} q^{\tilde{n}(\tilde{n}-1)} \sum_{j_l=1}^{\tilde{n}-1} (-1)^{j_l}= \frac{1}{2}q^{\tilde{n}(\tilde{n}-1)}.
\end{align}
We accordingly obtain 
\begin{align}
\tau_N^\nu &\sim (-1)^{M(N-M)+\binom{M}{2}+\binom{N-M}{2}} \times\nonumber \\
&\prod_{k=1}^{N} q^{(n+3k-3)\nu} \prod_{k=M+1}^{N} q^{2(n+2k-2)(N-k)} \prod_{k=1}^{N-M-1}q^{2k(N-M-1-k)} \prod_{k=n+N-1}^{n+N+M-2} \frac{1}{2}q^{k(k-1)} \nonumber \\
&= \frac{(-1)^{\binom{N}{2}}}{2^{M}} q^{\frac{\nu N}{2} (2n+3N-3)+\frac{2}{3}\binom{N-M}{2}(3n+2N+4M-4)+2 \binom{N-M}{3}+\frac{1}{2}\binom{M+1}{3}+2M\binom{n+N+\frac{M-3}{2}}{2}} \nonumber \\
&= \frac{(-1)^{\binom{N}{2}}}{2^{M}} q^{\frac{\nu N}{2} (2n+3N-3)+2 \binom{N-M}{2}(n+N+M-2)+\frac{1}{2}\binom{M+1}{3}+2M\binom{n+N+\frac{M-3}{2}}{2}}. \label{eq:B_result2}
\end{align}
\subsection{Case (C-b)} \label{sec:caseCb}
In this case, all types of $\mathrm{J}_\nu(n)$'s appear as anti-diagonal elements in \eqref{eq:tau_det}. Although our discussion is similar to the case (B), elements of type \eqref{eq:ev_n<<-1} contribute to the dominant term. The result is represented as
\begin{align}
\tau_N^\nu \sim {}&(-1)^{\binom{N}{2}} q^{\frac{\nu N}{2} (2n+3N-3)+2 \binom{N-M}{2}(n+N+M-2)} \prod_{k=-\nu}^{n+N+M-2} \frac{1}{2}q^{k(k-1)} \nonumber \\
&\times \prod_{k=n+N-1}^{-\nu-1} \left\{(-1)^{\binom{k+\nu+1}{2}} p_2(k+\nu) q^{\varphi_\nu(k) +p_1(k+\nu)-\nu k}\right\}.
\end{align}
\section{Concluding Remarks}
We have given the ultradiscrete Bessel function by taking ultradiscrete limit of the $q$-Bessel function $J_\nu (q^n)$. The $q$-Euler transformation helps us to evaluate $J_\nu (q^n)$ for $n \le 0$. This technique appeared in \cite{IST}. Inconsistency between the ultradiscrete Bessel function and the solution of an initial value problem to the ultradiscrete Bessel equation has been clarified, which gives us a future problem. 

Based on this result, we have constructed special solutions for the ultradiscrete Painlev\'{e} III equation by ultradiscretizing those for the discrete Painlev\'{e} III equation represented by the Casorati determinants whose elements are given by the $q$-Bessel function. The p-ultradiscretization enables us to take ultradiscrete limit of the determinants. Although evaluation for determinants has discussed in \cite{IST}, calculation in subsection \ref{sec:caseB} and \ref{sec:caseCb} is newly studied. It is a future problem to study ultradiscrete analogs of other Painlev\'{e} equations and discuss their mathematical structure such as degeneration structure. 
\section*{Acknowledgment}
This work was supported by JSPS KAKENHI Grant Number 26790082. 
\section*{Appendix A}
The ultradiscrete Bessel function $\mathcal{B}_n^\nu=(\beta^\nu_n, B^\nu_n)$ for $\nu>0$ by Mr.~Narasaki \cite{Narasaki} is written as 
\begin{eqnarray}
\begin{array}{l}
 \beta^\nu_n=\left\{ \begin{array}{ll}
 \beta^\nu_0 &(n\geq -\nu) \\
 (-1)^{\frac{n+\nu}{2}}\beta^\nu_0 & (n\leq -\nu -1,\ n+\nu:{\rm even}) \\
 (-1)^{\frac{n+\nu -1}{2}}\beta^\nu_0 & (n\leq -\nu -1,\ n+\nu:{\rm odd})\\
 \end{array} \right. \\
 B^\nu_n=\left\{ \begin{array}{ll}
 B^\nu_0+n\nu Q &(n\geq 0) \\
 B^\nu_0+n(n+\nu-1)Q & (-1 \geq n \geq -\nu ) \\
 B^\nu_0+\frac{n(n-2)-\nu^2}{2}Q & (n\leq -\nu -1,\ n+\nu:{\rm even}) \\
 B^\nu_0+\frac{n(n-4)-(\nu +3)(\nu -1)}{2}Q & (n\leq -\nu -1,\ n+\nu:{\rm odd})\\
 \end{array} \right.   
 \end{array}   \label{eq:ud-bessel}
\end{eqnarray}
and that for $\nu=0$ is 
\begin{eqnarray}
\begin{array}{l}
 \beta^0_n=\left\{ \begin{array}{ll}
 \beta^0_0 & (n\geq 1) \\
 (-1)^{\frac{n}{2}}\beta^0_0 & (n\leq -1 :{\rm even}) \\
 (-1)^{\frac{n-1}{2}}\beta^0_0 & (n\leq -1 :{\rm odd}) \\
 \end{array} \right. \\[15pt]
 B^0_n=\left\{ \begin{array}{ll}
 B^0_0 & (n\geq 1) \\
 B^0_0+\frac{n(n-2)}{2}Q & (n\leq -1 :{\rm even}) \\
 B^0_0+\frac{(n-1)(n-3)}{2}Q & (n\leq -1 :{\rm odd}). \\
 \end{array} \right. 
\end{array}   \label{eq:ultra-bessel-3}
\end{eqnarray}
This solution is constructed as a solution of an initial value problem for \eqref{eq:max-bessel}. 
\section*{Appendix B} 
We present a procedure to deform $\tilde{\tau}_N^\nu(n)$ in subsection \ref{sec:caseB}. \\\textit{Procedure}:
\begin{itemize}
\item[step 0-1] Add the $N$th row $\times$ $(-1) $ to the first row.
\item[step 0-2] Add the $N$th row $\times$ $(-1) $ to the second row.\\
$\vdots$
\item[step 0-$M$] Add the $N$th row $\times$ $(-1) $ to the $M$th row.
\item[step 1-1] Add the ($N-1$)th column $\times$ $-q^{2(N-M-1)}$ to the $N$th column.
\item[step 1-2] Add the ($N-2$)th column $\times$ $-q^{2(N-M-1)}$ to the ($N-1$)th column.\\
$\vdots$
\item[step 1-($N-1$)] Add the first column $\times$ $-q^{2(N-M-1)}$ to the second column.
\item[step 1-Ex.] Expand the determinant with the ($M+1$)th row. We obtain the determinant of size $(N-1)\times(N-1)$.
\item[step 2-1] For the resulting determinant, add the ($N-2$)th column $\times$ $-q^{2(N-M-2)}$ to the ($N-1$)th column.
\item[step 2-2] Add the ($N-3$)th column $\times$ $-q^{2(N-M-2)}$ to the ($N-2$)th column.\\
$\vdots$
\item[step 2-($N-2$)] Add the first column $\times$ $-q^{2(N-M-2)}$ to the second column.
\item[step 2-Ex.] Expand the determinant with ($M+1$)th row. We obtain the minor of size $(N-2)\times(N-2)$.\\
$\vdots$
\item[step ($N-M$)-1] For the resulting determinant of size $(M+1)\times(M+1)$, add the $M$th column $\times$ $(-1)$ to the $(M+1)$th column.
\item[step ($N-M$)-2] Add the $M-1$th column $\times$ $(-1)$ to the $M$th column.\\
$\vdots$
\item[step ($N-M$)-$M$] Add the first column $\times$ $(-1)$ to the second column.
\item[step ($N-M$)-Ex.] Expand the determinant with ($M+1$)th row. We obtain the minor of size $M\times M$.
\end{itemize}

\end{document}